%Paper: gr-qc/9307013
%From: j_halliwell@icva.DNET.NASA.GOV
%Date: Mon, 12 Jul 93 12:07:03 -0400

%% This paper is written in plain Tex and should run as it is, with
%% no input macros. It contains jnl.tex and reforder.tex
%%
%%      JNL.TEX

\font\twelverm=cmr10 scaled 1200    \font\twelvei=cmmi10 scaled 1200
\font\twelvesy=cmsy10 scaled 1200   \font\twelveex=cmex10 scaled 1200
\font\twelvebf=cmbx10 scaled 1200   \font\twelvesl=cmsl10 scaled 1200
\font\twelvett=cmtt10 scaled 1200   \font\twelveit=cmti10 scaled 1200
\font\twelvesc=cmcsc10 scaled 1200  %\font\twelvesf=cmssmc10 scaled 1200
\skewchar\twelvei='177   \skewchar\twelvesy='60

\def\twelvepoint{\normalbaselineskip=12.4pt plus 0.1pt minus 0.1pt
  \abovedisplayskip 12.4pt plus 3pt minus 9pt
  \belowdisplayskip 12.4pt plus 3pt minus 9pt
  \abovedisplayshortskip 0pt plus 3pt
  \belowdisplayshortskip 7.2pt plus 3pt minus 4pt
  \smallskipamount=3.6pt plus1.2pt minus1.2pt
  \medskipamount=7.2pt plus2.4pt minus2.4pt
  \bigskipamount=14.4pt plus4.8pt minus4.8pt
  \def\rm{\fam0\twelverm}          \def\it{\fam\itfam\twelveit}%
  \def\sl{\fam\slfam\twelvesl}     \def\bf{\fam\bffam\twelvebf}%
  \def\mit{\fam 1}                 \def\cal{\fam 2}%
  \def\sc{\twelvesc}               \def\tt{\twelvett}
  \def\sf{\twelvesf}
  \textfont0=\twelverm   \scriptfont0=\tenrm   \scriptscriptfont0=\sevenrm
  \textfont1=\twelvei    \scriptfont1=\teni    \scriptscriptfont1=\seveni
  \textfont2=\twelvesy   \scriptfont2=\tensy   \scriptscriptfont2=\sevensy
  \textfont3=\twelveex   \scriptfont3=\twelveex  \scriptscriptfont3=\twelveex
  \textfont\itfam=\twelveit
  \textfont\slfam=\twelvesl
  \textfont\bffam=\twelvebf \scriptfont\bffam=\tenbf
  \scriptscriptfont\bffam=\sevenbf
  \normalbaselines\rm}

%       tenpoint

\def\beginlinemode{\endmode
  \begingroup\parskip=0pt \obeylines\def\\{\par}\def\endmode{\par\endgroup}}
\def\beginparmode{\endmode
  \begingroup \def\endmode{\par\endgroup}}
\let\endmode=\par
{\obeylines\gdef\
{}}
\def\singlespace{\baselineskip=\normalbaselineskip}

\def\oneandahalfspace{\baselineskip=\normalbaselineskip
  \multiply\baselineskip by 3 \divide\baselineskip by 2}
\def\doublespace{\baselineskip=\normalbaselineskip \multiply\baselineskip by 2}

\newcount\firstpageno
\firstpageno=2
\footline={\ifnum\pageno<\firstpageno{\hfil}\else{\hfil\twelverm\folio\hfil}\fi}
\def\toppageno{\global\footline={\hfil}\global\headline
  ={\ifnum\pageno<\firstpageno{\hfil}\else{\hfil\twelverm\folio\hfil}\fi}}
\let\rawfootnote=\footnote              % We must set the footnote style
\def\footnote#1#2{{\rm\singlespace\parindent=0pt\parskip=0pt
  \rawfootnote{#1}{#2\hfill\vrule height 0pt depth 6pt width 0pt}}}
\def\raggedcenter{\leftskip=4em plus 12em \rightskip=\leftskip
  \parindent=0pt \parfillskip=0pt \spaceskip=.3333em \xspaceskip=.5em
  \pretolerance=9999 \tolerance=9999
  \hyphenpenalty=9999 \exhyphenpenalty=9999 }
\def\dateline{\rightline{\ifcase\month\or
  January\or February\or March\or April\or May\or June\or
  July\or August\or September\or October\or November\or December\fi
  \space\number\year}}
\def\received{\vskip 3pt plus 0.2fill
 \centerline{\sl (Received\space\ifcase\month\or
  January\or February\or March\or April\or May\or June\or
  July\or August\or September\or October\or November\or December\fi
  \qquad, \number\year)}}

\hsize=6.5truein
%\hoffset=1truein
\vsize=8.9truein
%\voffset=1truein
\parskip=\medskipamount
\def\\{\cr}
\twelvepoint            % selects twelvepoint fonts (cf. \tenpoint)
\doublespace            % selects double spacing for main part of paper (cf.
                        %       \singlespace, \oneandahalfspace)
\overfullrule=0pt       % delete the nasty little black boxes for overfull box

\def\title                      %  Title on title page
  {\null\vskip 3pt plus 0.2fill
   \beginlinemode \doublespace \raggedcenter \bf}

\def\author                     %  Author(s) name(s)  on title page
  {\vskip 3pt plus 0.2fill \beginlinemode
   \singlespace \raggedcenter\sc}

\def\affil                      % Affiliations (can intermix with \author)
  {\vskip 3pt plus 0.1fill \beginlinemode
   \oneandahalfspace \raggedcenter \sl}

\def\abstract                   % Begin abstract
  {\vskip 3pt plus 0.3fill \beginparmode
   \singlespace ABSTRACT: }

\def\endtopmatter               % End title page, begin body of paper
  {\endpage                     %       This subsumes \body
   \body}

\def\body                       % Begin text body;  can be used to end
  {\beginparmode}               % \title, \author, \affil, \abstract,
                                % \reference, or \figurecaption modes

\def\head#1{                    % Head;  NOTE enclose the text in {}
  \goodbreak\vskip 0.5truein    %  e.g., \head{I. Introduction}
  {\immediate\write16{#1}
   \raggedcenter \uppercase{#1}\par}
   \nobreak\vskip 0.25truein\nobreak}

\def\subhead#1{                 % Subhead;  NOTE enclose the text in {}
  \vskip 0.25truein             % e.g., \subhead{A. History of the Problem}
  {\raggedcenter {#1} \par}
   \nobreak\vskip 0.25truein\nobreak}

\def\beginitems{
\par\medskip\bgroup\def\i##1 {\item{##1}}\def\ii##1 {\itemitem{##1}}
\leftskip=36pt\parskip=0pt}
\def\enditems{\par\egroup}

\def\beneathrel#1\under#2{\mathrel{\mathop{#2}\limits_{#1}}}

\def\refto#1{$^{#1}$}           % For references in text as superscript

\def\references                 % Begin references -- basic format is Phys Rev
  {\head{References}            % I.e., volume, page, year (space after
%%commas).
   \beginparmode
   \frenchspacing \parindent=0pt \leftskip=1truecm
   \parskip=8pt plus 3pt \everypar{\hangindent=\parindent}}

\gdef\refis#1{\item{#1.\ }}                     % Ref list numbers.

\gdef\journal#1, #2, #3, 1#4#5#6{               % Journal reference.  Comma
%%sets
    {\sl #1~}{\bf #2}, #3 (1#4#5#6)}            % off: name, vol, page, year

\gdef\refa#1, #2, #3, #4, 1#5#6#7.{\noindent#1, #2 {\bf #3}, #4 (1#5#6#7).\rm}
%refa: type in: name,
%journal, vol, page, year
%prints out in same order

\gdef\refb#1, #2, #3, #4, 1#5#6#7.{\noindent#1 (1#5#6#7), #2 {\bf #3}, #4.\rm}
%refb: reads in same
%prints out name (year) etc.

\def\endreferences{\body}

\def\figurecaptions             % Begin figure captions
  {\endpage
   \beginparmode
   \head{Figure Captions}
}

\def\endpage                    %  Eject a page
  {\vfill\eject}

\def\endpaper                   %  Ways to say goodbye
  {\endmode\vfill\supereject}

\def\heading                            % Heading
  {\vskip 0.5truein plus 0.1truein      % e.g., \heading I. NOTES \endheading
   \beginparmode \def\\{\par} \parskip=0pt \singlespace \raggedcenter}

\def\subheading                         % Subheading
  {\vskip 0.25truein plus 0.1truein     % e.g., \subheading{A. The Problem}
   \beginlinemode \singlespace \parskip=0pt \def\\{\par}\raggedcenter}

\def\tag#1$${\eqno(#1)$$}

\def\align#1$${\eqalign{#1}$$}

\def\aligntag#1$${\gdef\tag##1\\{&(##1)\cr}\eqalignno{#1\\}$$
  \gdef\tag##1$${\eqno(##1)$$}}

\def\overset #1\to#2{{\mathop{#2}\limits^{#1}}}
\def\underset#1\to#2{{\let\next=#1\mathpalette\undersetpalette#2}}
\def\undersetpalette#1#2{\vtop{\baselineskip0pt
\ialign{$\mathsurround=0pt #1\hfil##\hfil$\crcr#2\crcr\next\crcr}}}

%%
%%      Various little user definitions
%%

\def\ref#1{Ref.~#1}                     %       for inline references
\def\Ref#1{Ref.~#1}                     %       ditto
\def\[#1]{[\cite{#1}]}
\def\cite#1{{#1}}
%%\def\Equation#1{Equation~(#1)}                % For citation of equation
%%numbe
%%\def\Equations#1{Equations~(#1)}      %       ditto
%%\def\Eq#1{Eq.~(#1)}                   %       ditto
%%\def\Eqs#1{Eqs.~(#1)}                 %       ditto
\def\(#1){(\call{#1})}
\def\call#1{{#1}}
\def\taghead#1{}
\def\frac#1#2{{#1 \over #2}}
\def\half{{\frac 12}}

\def\12{{1\over2}}

\def\ie{{\it i.e.}}

\def\etc{{\it etc.}}

\def\sla{\raise.15ex\hbox{$/$}\kern-.57em}
\def\leaderfill{\leaders\hbox to 1em{\hss.\hss}\hfill}
\def\twiddle{\lower.9ex\rlap{$\kern-.1em\scriptstyle\sim$}}
\def\bigtwiddle{\lower1.ex\rlap{$\sim$}}
\def\gtwid{\mathrel{\raise.3ex\hbox{$>$\kern-.75em\lower1ex\hbox{$\sim$}}}}
\def\ltwid{\mathrel{\raise.3ex\hbox{$<$\kern-.75em\lower1ex\hbox{$\sim$}}}}
\def\square{\kern1pt\vbox{\hrule height 1.2pt\hbox{\vrule width 1.2pt\hskip 3pt
   \vbox{\vskip 6pt}\hskip 3pt\vrule width 0.6pt}\hrule height 0.6pt}\kern1pt}
\def\tdot#1{\mathord{\mathop{#1}\limits^{\kern2pt\ldots}}}

\def\pmb#1{\setbox0=\hbox{#1}%
  \kern-.025em\copy0\kern-\wd0
  \kern  .05em\copy0\kern-\wd0
  \kern-.025em\raise.0433em\box0 }

%%%% definitions

\def\la{\langle}
\def\ra{\rangle}
\def\ria{\rightarrow}
\def\x{{\bar x}}
\def\k{{\bar k}}
\def\F{{\bar F}}
\def\a{\alpha}

\def\U{\Upsilon}

\def\Tr{{\rm Tr}}
\def\ih{{ {i \over \hbar} }}
\def\au{{\underline{\alpha}}}
\def\s{{\sigma}}
\def\ts{{\tilde \sigma}}

\def\pp{{\prime\prime}}
\def\xpp{{x^{\pp}}}
\def\S{{\cal S}}
\def\A{{\cal A}}
\def\E{{\cal E}}

%%		REFORDER.TEX			6/7/85	Doug E.

\catcode`@=11
\newcount\r@fcount \r@fcount=0
\newcount\r@fcurr
\immediate\newwrite\reffile
\newif\ifr@ffile\r@ffilefalse
\def\w@rnwrite#1{\ifr@ffile\immediate\write\reffile{#1}\fi\message{#1}}

\def\writer@f#1>>{}
\def\referencefile{%			  Stuff to write .REF file
  \r@ffiletrue\immediate\openout\reffile=\jobname.ref%
  \def\writer@f##1>>{\ifr@ffile\immediate\write\reffile%
    {\noexpand\refis{##1} = \csname r@fnum##1\endcsname = %
     \expandafter\expandafter\expandafter\strip@t\expandafter%
     \meaning\csname r@ftext\csname r@fnum##1\endcsname\endcsname}\fi}%
  \def\strip@t##1>>{}}

\def\citeall#1{\xdef#1##1{#1{\noexpand\cite{##1}}}}
\def\cite#1{\each@rg\citer@nge{#1}}	% Variable No. of args, separated by

\def\each@rg#1#2{{\let\thecsname=#1\expandafter\first@rg#2,\end,}}
\def\first@rg#1,{\thecsname{#1}\apply@rg}	% each@ag is a general purpose
\def\apply@rg#1,{\ifx\end#1\let\next=\relax%	  variable no. of arg. macro.
\else,\thecsname{#1}\let\next=\apply@rg\fi\next}% args separated by commas

\def\citer@nge#1{\citedor@nge#1-\end-}	% Check for M-N range (M and N numbers)
\def\citer@ngeat#1\end-{#1}
\def\citedor@nge#1-#2-{\ifx\end#2\r@featspace#1 % Single argument
  \else\citel@@p{#1}{#2}\citer@ngeat\fi}	% M-N range of arguments
\def\citel@@p#1#2{\ifnum#1>#2{\errmessage{Reference range #1-#2\space is bad.}%
    \errhelp{If you cite a series of references by the notation M-N, then M and
    N must be integers, and N must be greater than or equal to M.}}\else%
 {\count0=#1\count1=#2\advance\count1 by1\relax\expandafter\r@fcite\the\count0,
  \loop\advance\count0 by1\relax%	  Loop from M to N
    \ifnum\count0<\count1,\expandafter\r@fcite\the\count0,%
  \repeat}\fi}

\def\r@featspace#1#2 {\r@fcite#1#2,}	% Eat spaces at beginning or end of arg
\def\r@fcite#1,{\ifuncit@d{#1}%		  Cite individual reference
    \newr@f{#1}%
    \expandafter\gdef\csname r@ftext\number\r@fcount\endcsname%
                     {\message{Reference #1 to be supplied.}%
                      \writer@f#1>>#1 to be supplied.\par}%
 \fi%
 \csname r@fnum#1\endcsname}
\def\ifuncit@d#1{\expandafter\ifx\csname r@fnum#1\endcsname\relax}%
\def\newr@f#1{\global\advance\r@fcount by1%
    \expandafter\xdef\csname r@fnum#1\endcsname{\number\r@fcount}}

\let\r@fis=\refis			% Save old \refis, redefine
\def\refis#1#2#3\par{\ifuncit@d{#1}%      Use two params #2 #3 to strip blank
   \newr@f{#1}%
   \w@rnwrite{Reference #1=\number\r@fcount\space is not cited up to now.}\fi%
  \expandafter\gdef\csname r@ftext\csname r@fnum#1\endcsname\endcsname%
  {\writer@f#1>>#2#3\par}}

\def\ignoreuncited{%   redefine \refis if ignoring uncited references
   \def\refis##1##2##3\par{\ifuncit@d{##1}%
    \else\expandafter\gdef\csname r@ftext\csname r@fnum##1\endcsname\endcsname%
     {\writer@f##1>>##2##3\par}\fi}}

\def\r@ferr{\endreferences\errmessage{I was expecting to see
\noexpand\endreferences before now;  I have inserted it here.}}
\let\r@ferences=\references
\def\references{\r@ferences\def\endmode{\r@ferr\par\endgroup}}

\let\endr@ferences=\endreferences
\def\endreferences{\r@fcurr=0%		  Save old \endreferences, redefine
  {\loop\ifnum\r@fcurr<\r@fcount%	  Loop over refnum and produce text
    \advance\r@fcurr by 1\relax\expandafter\r@fis\expandafter{\number\r@fcurr}%
    \csname r@ftext\number\r@fcurr\endcsname%
  \repeat}\gdef\r@ferr{}\endr@ferences}

% Save old \endpaper, redefine it to write parting message.

\let\r@fend=\endpaper\gdef\endpaper{\ifr@ffile
\immediate\write16{Cross References written on []\jobname.REF.}\fi\r@fend}

\catcode`@=12

\citeall\refto		% These macros will generate citations
\citeall\ref		%
\citeall\Ref		%

%%%% Paper begins here.

\centerline{\bf Quantum-Mechanical Histories and the
Uncertainty Principle. }

\centerline{\bf II. Fluctuations about Classical Predictability}

\vskip 0.5in
\author J. J. Halliwell
\vskip 0.2in
\affil
Theory Group
Blackett Laboratory
Imperial College
London SW7 2BZ
UK
\vskip 0.5in
\centerline {\rm Preprint IC 92-93/XX, July, 1993}
\vskip 0.2in
\centerline {\rm Submitted to {\sl Physical Review D}}.
\endpage
\abstract
{This paper is concerned with two questions in the decoherent
histories approach to quantum mechanics: the emergence of
approximate classical predictability, and the fluctuations about it
necessitated by the uncertainty principle. It is in part a
continuation of an earlier paper on the uncertainty principle for
quantum-mechanical histories.
We consider histories
characterized by position samplings at $n$ moments of time. This
includes the case of position sampling continuous in time in the
limit $n \rightarrow \infty $.  We use this to construct a
probability distribution on the value of (discrete approximations
to) the field equations, $F = m \ddot x + V'(x) $, at $n-2$ times.
We find that it is peaked around $F=0$; thus classical
correlations are exhibited. We show that the  width
of the peak $ \Delta F$
is largely independent of the initial state.
We show that the uncertainty principle takes the form
$2 \sigma^2  \ (\Delta F)^2 \ge { \hbar^2  / t^2 } $,
where $\sigma$ is the  width of
the position samplings, and $t$ is the timescale between
projections. We determine the modifications to this result when the
system is coupled to a thermal environment, thus obtaining a measure
of the comparative sizes of quantum and thermal fluctuations. We
show that the thermal fluctuations become comparable with the
quantum fluctuations under the same conditions that decoherence
effects come into play, in argreement with earlier work.
We also study an alternative measure of
classical correlations, namely the conditional probability of
finding a sequence of position samplings, given that particular
initial phase space data have occurred. We use these results to
address the issue of the formal interpretation  of the
probabilities for sequences of position samplings in the decoherent
histories approach to quantum mechanics. Under appropriate
conditions, which we describe, they admit an interpretation as a
statistical ensemble of classical solutions, with the probability of
each individual classical solution given by a smeared Wigner
function of its initial data.
We study the decoherence properties of histories characterized by
the value of the field equations, $F$, at a sequences of times. We
argue that they will be decoherent if their initial data are
decoherent.}

\endtopmatter

\head {\bf I. Introduction}

There has been considerable recent interest in the decoherent
histories approach to the quantum mechanics of closed systems
[\cite{gell,gell2,hartle,griffiths,omnes,dowker,albrecht,blencowe,
brun,calzetta,diosi,isham,twamley,saunders}].
Central to that
approach is the formula for the probability of a quantum-mechanical
history:
$$
p(\a_1, \a_2, \cdots \a_n) = \Tr \left( P_{\a_n}^n(t_n)\cdots
P_{\a_1}^1(t_1)
\rho P_{\a_1}^1 (t_1) \cdots P_{\a_n}^n (t_n) \right)
\eqno(1.1)
$$
Here, $\rho$ is the initial density matrix of the system, and the
histories of the system are characterized by strings of
projection operators $P_{\a_1}^1 \cdots P_{\a_n}^n$ at times $t_1
\cdots t_n$. The projections effect a partition of the possible
alternatives $\a_k$ of the system at time $t_k$. They are positive
hermitian operators that are both exclusive and exhaustive:
$$
\eqalignno{
P_{\a} P_{\beta} & = \delta_{\a \beta} \ P_{\a},
&(1.2) \cr
\sum_{\a} P_{\a} & = 1
&(1.3) \cr }
$$
Evolution between each projection is described by the unitary evolution
operator, $e^{-\ih Ht}$, and the time-dependent projections
$P_{\a_k}^k(t)$ appearing in (1.1) are given in terms of the
time-independent ones $P_{\a_k}^k $ by
$$
P^k_{\a_k}(t_k) = e^{\ih H(t_k-t_0)} P^k_{\a_k} e^{-\ih H(t_k-t_0)}
\eqno(1.4)
$$

Not all sets of histories may be assigned probabilities.
Interference is generally an obstruction to the probabilities
satisfying the Kolmogorov axioms, and in particular, to the requirement
that they be additive on disjoint regions of sample space. The
determination of which sets may be assigned probabilities is carried
out using the decoherence functional:
$$
D({\underline {\a}} , {\underline {\a}'} ) =
\Tr \left( P_{\a_n}^n(t_n)\cdots
P_{\a_1}^1(t_1)
\rho P_{\a_1'}^1 (t_1) \cdots P_{\a_n'}^n (t_n) \right)
\eqno(1.5)
$$
Here $ {\underline {\a}} $ denotes the string $\a_1, \a_2, \cdots
\a_n$. If the decoherence functional vanishes for all distinct pairs of
histories ${\underline {\a}}, {\underline {\a}'}$, then there is no
interference between them and probabilities may be assigned.
Sets of such histories are said to be consistent, or decoherent.

We are concerned in this paper with the question of how the
uncertainty principle arises in formulations of quantum mechanics
based on Eq.(1.1). This investigation was commenced in a previous
paper [\cite{jjh2}]. There it was argued that the uncertainty
principle appears as a lower bound on the Shannon information of
(1.1), and is most generally of the form,
$$
\eqalignno{
I & = \ - \sum_{\a_1 \cdots \a_n}  p(\a_1, \cdots \a_n) \ln
p(\a_1 \cdots \a_n )
\cr &
\ge \ \ln \left( { V_H \over
\s_1 \s_2^2 \cdots \s_{n-1}^2 \s_n } \right)
&(1.6) \cr }
$$
In Eq.(1.6), $\s_k$ is the width of the projection $P^k_{\a_k}$
onto any continuous variable at time $t_k$,
and $V_H$ is an elementary volume of history space, defined in
Ref.[\cite{jjh2}].
For example,
for a history characterized by a position projection
and a momentum projection immediately afterwards, $V_H = 2 \pi \hbar$.
For more general histories, it is given in terms of the short time
limit of certain
propagator prefactors.

A notion as basic as the uncertainty principle, however, must be
expected to arise in many different guises.
In the present paper, we shall obtain a rather different expression
of the uncertainty principle for the particular but important case of
histories characterized by
imprecise samplings of position at a sequence of times, including
the case of sampling continuous in time. On general grounds, one
would expect that successive position samplings will be strongly
correlated according to classical laws.
Our first task is to demonstrate this.
There are a variety of different
ways in which the notion of peaking about a classical history may be
expressed. A particular method we shall focus on is to replace the $n$
sampled positions $\x_1, \x_2  \cdots \x_n$  with the initial data
$ \x_1$, $\k_1$ together with (discrete approximations to) the value
of the field equations $F = m \ddot x + V'(x) $ at the remaining
$(n-2)$ moments of time. That is, in the case of samplings at equal
time separations $t$,
$$
\eqalignno{
\k_1 &= m {(\x_2 -\x_1) \over t}
&(1.7) \cr
\F_i &= m { (\x_{i+1} - 2 \x_i + \x_{i-1} ) \over t^2 } + V'(\x_i),
\quad \quad i = 2 ,3 \cdots (n-1).
&(1.8) \cr }
$$
By integrating out
the initial data $\x_1, \k_1 $
one obtains a probability distribution on the value
of the field equations, $ p(\F_2, \cdots \F_{n-1}) $. We shall
demonstrate that this probability distribution is largely
independent of the initial state, $\rho$, and furthermore, it is
peaked at $\F_i=0$. Classical correlations are thus exhibited.

The role of the uncertainty principle is to impose a
restriction on the width $\Delta F$ of the peak about $F=0$. In
particular, we shall show that this restriction generally takes the
form
$$
\s^2 \ (\Delta F)^2 \ge { \hbar^2 \over 2 t^2}
\eqno(1.9)
$$
where $\s$ is the width of the position samplings. A similar
result is derived for samplings continuous in time. All of the above
is discussed in Section 2, for the case of the free particle, and
in Section 3 for general potentials.

More germane to the decoherence programme is the question of
modifications of the uncertainty principle due to thermal
fluctuations. Many models of decoherence involve coupling a
distinguished system to a thermal environment. Under such
circumstances, one would expect the width of the peak about
classical histories to be broadened by thermal fluctuations, and
thus (1.9) will be modified. In Section 4, we therefore discuss the
form these modifications take. The question of thermal modifications
for information-theoretic measures of uncertainty has been discussed
in Ref.[\cite{ander}]. Here, we shall compute the width of the peak
about $F=0$ in the presence of thermal fluctuations. We thus
determine the regime in which thermal fluctutations become important
in comparison to quantum fluctuations. We find this regime to
coincide with the regime in which decoherence effects come into play.
We also discuss the connection
of this part of our work with that of Gell-Mann and Hartle
[\cite{gell2}].

As stated, there are a number of ways of exhibiting classical
correlations in a sequence of position sampling. In Section 5, we
discuss an alternative method. This is to consider the conditional
probability distribution
$$
p(\x_3, \cdots \x_n| \x_1, \k_1) =
{ p(\x_1, \k_1, \x_3, \cdots \x_n) \over p(\x_1, \k_1) }
\eqno(1.10)
$$
This is the probability of the sequence $\x_3 \cdots
\x_n$, given that $\x_1$, $\k_1$ have occurred, where we have used
(1.7) to replace $\x_2$ with the initial momentum $\k_1$.
Classical correlations are exhibited if (1.10) is peaked
about the classical solution with initial data $\x_1$, $\k_1$. We
shall show that this does indeed occur when the distribution
$p(\F_2, \cdots \F_{n-1} )$ is peaked around $\F_i = 0$. We use this
result to discuss the general interpretation of the probabilities for
strings of position samplings.

The considerations of Section 2 to 5 are largely concerned with the
properties of the expression for the probabilities, Eq.(1.1), for
the case of position samplings. They are not concerned with how the
decoherence condition is satisfied. This will be discussed in
Section 6. It will be argued that diagonality of the decoherence
functional in the initial data, $\x_1, \k_1$, together with the
classical predictability indicated by the peak about $\F_i=0$, are
sufficient to ensure the consistency of histories of $\F_i$'s.
The discussion of decoherence perhaps falls logically prior to that
of classical correlations, in Section 2--5, but it is in some ways
easier to present the discussion in this order, since some of the
discussion in Section 6 relies on the results of earlier part of
the paper. We summarize and conclude in Section 7.

Classical predictability has of course been extensively discussed in
the literature on the decoherent histories approach
[\cite{gell2,dowker,brun,omnes2}]. What this paper has to add to the
discussion is a new method of exhibiting it. The approach is in some
ways more systematic and precise than, for example, that of
Refs.[\cite{gell2,dowker}], but the conclusions are the same. By contrast,
the question of the form of the uncertainty principle for
quantum-mechanical histories has received very little attention in
the literature\footnote{$^{\dag}$}{See, however,
Ref.[\cite{hartle2}] for some steps in this direction in the context
of the decoherent histories approach. Also, the question of the
uncertainty principle for position measurements continuous in time
has been considered by Mensky [\cite{mensky1}].},
and the main new results
presented in this paper consist of a study of this question.

Although aimed at the decoherent histories approach to quantum
mechanics, the results of this paper are not restricted to that
approach. As will be explained in Section 6, the formula (1.1) may
also be thought of as the probability for a sequence of
measurements in the Copenhagen approach to the quantum mechanics of
measured subsystems. Sections 2--5 are concerned with the
mathematical properties of (1.1), and are largely independent of which
interpretational scheme one wishes to adopt.

\head {\bf II. Peaking about Classical Paths}

The probability for a history consisting of $n$ position samplings is
$$
p(\x_1, \x_2, \cdots \x_n)
= \Tr \left[ P_{\x_n} e^{-iHt} \cdots
P_{\x_2} e^{-iHt} P_{\x_1} \rho P_{\x_1} e^{iHt} P_{\x_2} \cdots
P_{\x_n} \right]
\eqno(2.1)
$$
We take the samplings to be a time $t$ apart, and the projections
have the form
$$
P_{\x} = \int dx \ \U(x - \x) \ | x \ra \la x |
\eqno(2.2)
$$
where $\U$ is a sampling function\footnote{$^{*}$}{See
Ref.[\cite{caves}] for a thorough discussion of quantum mechanics
for samplings distributed in time.}.
The most appropriate choice for
$\U$ is to take it to be a function which vanishes outside an
interval of size $\s$, such as
$$
\U(x-\x)= \theta \left({x-\x + \half \sigma  \over \s } \right) \
\theta \left( {-x + \x + \half \sigma \over \s }\right)
\eqno(2.3)
$$
In this case, $\x$ is a discrete label. In this paper, we shall
largely concentrate on the technically easier case of a Gaussian
$$
\U(x-\x) = {1 \over ( 2 \pi \s^2 )^{\half} }
\ \exp \left( - { (x-\x)^2 \over 2 \sigma^2 } \right)
\eqno(2.4)
$$
Here $\x$ is now a continuous label and the projections are
only approximately exclusive. The probability (2.1) is a
probability density, not a probability, even in the discrete case
(2.3). (The notation is therefore different to that used in
Ref.[\cite{jjh2}]).

Given the distribution $p(\x_1, \x_2, \cdots \x_n)$, it turns out
to be convenient to perform the change of variables
$$
\eqalignno{
\k_1 &= m {(\x_2 -\x_1) \over t }
&(2.5) \cr
\F_i &= m {(\x_{i+1} - 2 \x_i + \x_{i-1} ) \over t^2}  + V'(\x_i),
\quad i = 2, \cdots n-1.
&(2.6) \cr }
$$
as indicated in the Introduction.
Thus $\k_1$ is a discrete approximation to the initial momentum and
$\F_2, \cdots \F_{n-1}$ are discrete approximations to the field
equations at $(n-2)$ times. We now focus on the probability
distribution
$ p (\x_1, \k_1, \F_2, \cdots F_{n-1}) $, a distribution on the initial
data together with the value of the field equations at $(n-2)$ times.
If $\F_2 \cdots \F_{n-1} $ are integrated out, a phase space
distribution is obtained. The expression of the uncertainty
principle for distributions of this type is discussed in
Refs.[\cite{jjh2,ander}]
(see also the comments in Section 5).
On the other hand, if the initial data $\x_1,\k_1$ are
integrated out, one obtains a distribution on the value of the field
equations,
$$
p(\F_2, \cdots \F_{n-1}) = \int d\x_1 d \k_1 \ p(\x_1, \k_1, \F_2,
\cdots \F_{n-1})
\eqno(2.7)
$$
We shall study this expression. We shall show, first of all, that
(2.7) is typically peaked about $\F_i = 0$, as one would expect.
Secondly, we shall show that the width $\Delta F$ of the peak is largely
independent of the initial state. We shall compute the width
explicitly, and show that the uncertainty principle arises in the
form,
$$
\s^2 (\Delta F)^2  \ \ge \ { \hbar^2 \over 2 t^2}
\eqno(2.8)
$$
where $\s$ is the width of the position projections and $t$ is the
time scale between the projections.

\subhead{\bf II(A). The Free Particle}

Now we compute the probabilities $p(\x_1, \cdots \x_n)$
for the free particle.
Explicitly, Eq.(2.1) is
$$
\eqalignno{
p(\x_1, \x_2, \cdots \x_n)
=& \int \ dx_n dy_n
\ \prod_{k=1}^{n-1} dx_k dy_k \ \delta(x_n - y_n)
\ J(x_{k+1}, y_{k+1}, t | x_k, y_k, 0 )
\cr & \times
\prod_{k=1}^n \ \U(x_k - \x_k) \U(y_k -\x_k) \ \rho(x_1, y_1)
&(2.9) \cr }
$$
where
$J$ is the density matrix propagator. Introduce $X= \half(x+y)$,
$\xi = x-y$. For the unitary free particle case $J$ is then given by
$$
J(x_{k+1}, y_{k+1}, t | x_k, y_k, 0 ) = { m \over  \pi \hbar t}
\ \exp \left( { i m \over \hbar t } (X_{k+1} - X_k) ( \xi_{k+1} -
\xi_k ) \right)
\eqno(2.10)
$$
It is convenient to write the initial density matrix $\rho(x_1,y_1)$
in terms of its Wigner representation [\cite{wignerfn}],
$$
\rho(x_1,y_1) = \rho(X_1 + \half \xi_1, X_1 - \half \xi_1) =
\int dp \ e^{\ih p \xi_1} \ W_\rho(p, X_1)
\eqno(2.11)
$$
The probability (2.9) thus becomes,
$$
\eqalignno{
p(\x_1, \x_2, \cdots \x_n)   =&  \int \prod_{k=1}^n dX_k d\xi_k \ dp
\ \delta(\xi_n) \ e^{\ih p \xi_1} \ W_{\rho}(p, X_1)
\cr & \times \exp \left( {im \over \hbar t} \sum_{k=1}^{n-1}
(X_{k+1} -X_k)( \xi_{k+1} - \xi_k ) \right)
\times \left( { m \over \pi \hbar t} \right)^{n-1}
\cr & \times   \prod_{k=1}^n \ \U (X_k + \half \xi_k - \x_k )
\ \U ( X_k - \half \xi_k - \x_k )
&(2.12) \cr }
$$
Now performing the shifts $X_k \rightarrow X_k + \x_k $,
$p \rightarrow p + m { (\x_2 - \x_1) \over t } $, and introducing
$\k_1$ and $\F_k$, defined above, Eq.(2.12) becomes,
$$
\eqalignno{
p(\x_1, \x_2, \cdots \x_n)   =&  \int \prod_{k=1}^n dX_k d\xi_k \ dp
\ \delta(\xi_n) \ W_{\rho}(p + \k_1 , X_1+ \x_1 )
\times \left( { m \over \pi \hbar t} \right)^{n-1}
\cr & \times  \exp \left( {im \over \hbar t} \sum_{k=1}^{n-1}
(X_{k+1} -X_k)( \xi_{k+1} - \xi_k ) - { i t \over \hbar}
\sum_{k=2}^{n-1} \F_k \xi_k + \ih p \xi_1  \right)
\cr & \times  \prod_{k=1}^n \ \U (X_k + \half \xi_k  )
\ \U ( X_k - \half \xi_k  )
\cr = &
\ N \ p(\x_1, \k_1, \F_2 \cdots \F_{n-1})
&(2.13) \cr }
$$
where $N$ is a Jacobean factor, which we do not give explicitly.

\subhead{\bf II(B). Approximate Projectors}

At this stage the sampling functions $\U$ are general -- they may
be exact or Gaussians. For simplicity we now work with
Gaussian samplings, and so the labels $\x$ are continuous.
As indicated at the beginning of this section, we may integrate out
the $\F_k$ to obtain a distribution on the initial data. This leads
to a distribution on the initial data given by a smeared Wigner
function. Its properties were worked out
Refs.[\cite{jjh}]. For us the more interesting thing is to integrate
out the initial data $\x_1$, $\k_1$. Carrying out the integration,
the Wigner function of the initial state drops out
entirely\footnote{$^{\dag}$}{Independence of the initial state in
this context has previously been noted by Caves [\cite{caves}].}.
The $p$ integration may be done, and the result is
$$
\eqalignno{
p(\F_2,  \cdots \F_{n-1})   =&  \int \prod_{k=1}^n dX_k d\xi_k
\ \delta(\xi_n) \ \delta (\xi_1)
\cr & \times  \exp \left( {im \over \hbar t} \sum_{k=1}^{n-1}
(X_{k+1} -X_k)( \xi_{k+1} - \xi_k ) - { i t \over \hbar}
\sum_{k=2}^{n-1} \F_k \xi_k \right)
\cr & \times  \prod_{k=1}^n \ \U (X_k + \half \xi_k  )
\ \U ( X_k - \half \xi_k  )
&(2.14) \cr }
$$
(hereafter we ignore prefactors -- they may be recovered in the
final answer using normalization conditions).
Eq.(2.14) is the probability distribution function for the field
equations $\F_k$. Note that the independence of the initial state
follows from the fact that the labels $\x$ are continous and
the Wigner function could be integrated out. This would not be
possible in the discrete case.

Consider now the evaluation of (2.14). Let us first consider
the case $n=3$, for which there is just one value of $\F_k$. The
integrals are straightforwardly carried out and the result is
$$
p( \F_2) = { 1 \over ( 2 \pi (\Delta F)^2 )^{\half} }
\ \exp \left( - { \F_2^2 \over 2 (\Delta F)^2} \right)
\eqno(2.15)
$$
where the width is
$$
(\Delta F)^2 = { \hbar^2 \over 2 \s^2 t^2 }
+ {4 m^2 \s^2 \over t^4}
\ \ge \ { \hbar^2 \over 2 \s^2 t^2 }
\eqno(2.16)
$$
Eq.(2.15) is peaked at $\F_2=0$. Its width satisfies the advertized
inequality (2.8).

Clearly the inequality (2.16) approaches equality when $ \s^2 <<
{\hbar t /  m} $. The significance of this regime was elucidated in
Ref.[\cite{jjh2}]. The quantity $m / \hbar t $ is the prefactor in the
propagator (2.9) and has the interpretation as a density of paths.
As discussed in Ref.[\cite{jjh2}],
its inverse $ \hbar t / m$
thus has the interpretation as
an ``elementary volume of history space'' (the factor $V_H$ in
Eq.(1.6)), analagous to the factor $2 \pi \hbar $ for samplings of
phase space. When the sampling volume $\s^2$ is much greater than
the elementary volume, the width (2.16) becomes independent of
$\hbar$, and the uncertainty is essentially classical uncertainty,
due to the imprecision of the sampling. On the other hand, the width
approaches the uncertainty principle limit (2.8) when the sampling
volume is less than the elementary history space volume.

More generally, we may evaluate (2.14) by taking the continuum
limit. This involves a few subtleties, so we describe it in some
detail. In (2.14), insert the Gaussian projections
explicitly, and also perform the discrete version of an integration
by parts. We thus obtain
$$
\eqalignno{
p(\F_2,  & \cdots  \F_{n-1})  =  \int \prod_{k=1}^n dX_k d\xi_k
\ \delta(\xi_n) \ \delta (\xi_1)
\cr & \times   \exp \left( - {im \over \hbar t} \sum_{k=2}^{n-1}
( \xi_{k+1} - 2 \xi_k + \xi_k)X_k  - { i t \over \hbar}
\sum_{k=2}^{n-1} \F_k \xi_k \right)
\cr \times \exp & \left( - \sum_{k=1}^n \left[ {X_k^2 \over \s_k^2}
+ { \xi_k^2 \over 4 \s_k^2} \right] + { im \over \hbar t}
\left[ X_n (\xi_n - \xi_{n-1}) - X_1 ( \xi_2 -\xi_1) \right] \right)
&(2.17) \cr }
$$
We have introduced a different width $\s_k$ for each moment of time.
Also, we have for notational convenience included $\xi_n$ and
$\xi_1$ explicitly, even though the delta-functions set them to zero.
Next we carry out the integrals over $X_n$ and $X_1$, obtaining,
$$
\eqalignno{
p(\F_2,  \cdots \F_{n-1}) &  =  \int \prod_{k=1}^n dX_k d\xi_k
\ \delta(\xi_n) \ \delta (\xi_1)
\cr & \times  \exp \left( - {im \over \hbar t} \sum_{k=2}^{n-1}
( \xi_{k+1} - 2 \xi_k + \xi_{k-1})X_k  - { i t \over \hbar}
\sum_{k=2}^{n-1} \F_k \xi_k \right)
\cr & \times  \exp \left( - \sum_{k=2}^{n-1} \left[ {X_k^2 \over \s_k^2}
+ { \xi_k^2 \over \s_k^2} \right] \right)
\cr & \times
\exp \left( - { m^2 \s_n^2 \over 4 \hbar^2 t^2 }  (\xi_n -\xi_{n-1})^2
- {m^2 \s_1^2 \over 4 \hbar^2 t^2 } (\xi_2 - \xi_1)^2
\right)
&(2.18) \cr }
$$

Now we take the continuum limit,
$t \ria 0 $, $ n \ria \infty $, with $ n t = T =
constant $. In order that the exponential in the Gaussian smearing
function go over to a well-defined continous integral,
it is also necessary to let $\s_k \ria \infty $ in the Gaussian
projections, and introduce a new width $ \tilde \s_k = t^{\half} \s_k $
which stays finite in the limit. The last exponentials in Eq.(2.18)
thus become delta-functions on the initial and final values of
$\dot \xi$, and one obtains the functional integral,
$$
\eqalignno{
p[ \F(t) ] = \int &  {\cal D} \xi {\cal D} X
\ \delta \left(\xi(0)\right)  \ \delta \left(\xi(T) \right)
\ \delta \left(\dot \xi(0)\right)  \ \delta \left(\dot \xi(T) \right)
\cr & \times
\exp \left( - { im \over \hbar } \int_0^T dt X \ \ddot \xi - \ih
\int_0^T dt  \F \xi \right)
\cr & \times
\exp \left( - \int_0^T dt \left[ { X^2 \over \tilde \s^2} + { \xi^2
\over 4 \tilde \s^2} \right] \right)
&(2.19) \cr }
$$
The boundary conditions on $\xi$ are those indicated by the
delta-functions; those on $X$ are that it is integrated over on
every time slice. Now we may perform the functional integration
over $X(t)$. The result is,
$$
\eqalignno{
p[ \F(t) ] = \int &  {\cal D} \xi
\ \delta \left(\xi(0)\right)  \ \delta \left(\xi(T) \right)
\ \delta \left(\dot \xi(0)\right)  \ \delta \left(\dot \xi(T) \right)
\cr & \times
\exp \left( - \int_0^T dt \ \xi
\left[ {m^2 \tilde \s^2 \over 4
\hbar^2} {d^4 \over dt^4 } + {1 \over 4 \tilde \s^2} \right]
\xi - \ih  \int_0^T dt  \F \xi \right)
&(2.20) \cr }
$$
where an integration by parts has been performed, and the boundary
conditions on $\xi$ invoked to drop the boundary terms.
Finally, the functional integration over $\xi$ may be perform, with
the formal result,
$$
p[ \F(t) ] = \exp \left( - { 1 \over 4 \hbar^2} \int dt dt' \F(t)
G(t,t') \F(t') \right)
\eqno(2.21)
$$
Here, $G(t,t')$ is the Green function of the fourth order operator
appearing in (2.20), {\it i.e.},
$$
\left[  {m^2 \tilde \s^2 \over 4
\hbar^2} {d^4 \over dt^4 } + {1 \over 4 \tilde \s^2} \right]
G(t,t') = \delta(t-t')
\eqno(2.22)
$$
with the boundary conditions that $G$ and its first derivative
vanish at both end points.

We shall not evaluate $G(t,t')$ exactly. However, it is clear that
for small $\ts$, one has
$$
G(t,t') \ \approx \ 4 \ts^2 \ \delta (t-t')
\eqno(2.23)
$$
and Eq.(2.21) may be written
$$
p[\F(t)] \ \approx \ \exp \left( - \int dt \ { \F^2(t)
\over 2 ( \Delta F)^2 } \right)
\eqno(2.24)
$$
where
$$
(\Delta F)^2
\ \approx \ { \hbar^2 \over  2 \ts^2 }
\eqno(2.25)
$$
It is also clear that (2.25) represents a lower bound
because, from (2.20)--(2.22), the neglected terms (involving the fourth
derivative) broaden the width.

Note that this differs slightly from Eq.(2.8). This is due to the
changing of the dimension of certain quantities in the continuum
limit: $\ts$ differs in dimension from $\s$ by
$({\rm time})^{\half}$,
and the natural definition of the width $\Delta F$ in (2.24)
differs from that of the discrete case by a
similar factor.

\subhead{\bf II(C). Exact Projectors}

To show that the results are not qualitatively different, we now
consider the case $n=3$ (as in Eqs.(2.15), (2.16)) with
exact projection operators, {\it i.e.}, the sampling functions are
given by (2.3). With exact projectors, the probabilities for
histories generally cannot be computed exactly. We will therefore
evaluate Eq.(2.13) in the interesting regime of small $\s$.

In the case of exact projectors, the labels $\x_k$ are discrete.
Integrating out the initial data we will therefore generally not
obtain the result (2.14), {\it i.e.}, the dependence on the initial
data will not drop out, and one is left with a factor of the form
$$
\sum_{\k_1, \x_1} \ W_{\rho}(p+\k_1,X_1+\x_1)
\eqno(2.26)
$$
In the approximate projector case, the labels $\k,\x$ are
continuous, and (2.26) is equal to $1$, using the normalization of
the Wigner function.
This will not be true in the
discrete case, and some dependence on the initial state will remain.
However, in the limit of small $\s$, the discrete sum
may be approximated by a continuous integral, (2.26) will then
be approximately equal to $1$, and Eq.(2.14) will approximately
hold, becoming exact in the limit $\s \ria 0 $.

The distribution of the value of the field equation, $\F_2$ is now
$$
\eqalignno{
p(\F_2) =  \left( { m \over \pi \hbar t} \right)^2
\int & dX_3 dX_2 dX_1 d\xi_2
\ \exp \left( - {im \over \hbar t} \xi_2 (X_3 -2X_2 +X_1) - {it
\over \hbar } \xi_2 \F_2 \right)
\cr &
\times \ \U(X_3) \ \U(X_2 + \half \xi_2)
\ \U(X_2 - \half \xi_2) \ \U(X_1)
&(2.27) \cr }
$$
The integrals over $X_1$ and $X_3$ are readily carried out with the
result,
$$
\eqalignno{
p(\F_2) = &   \left( { m \over \pi \hbar t} \right)^2
\int dX_2 d\xi_2
\ \exp \left( {2 im \over \hbar t} \xi_2 X_2  - {it
\over \hbar } \xi_2 \F_2 \right)
\cr &
\times \ \U(X_2 + \half \xi_2) \ \U(X_2 - \half \xi_2)
\ \left( {2 \hbar t \over m \xi_2} \right)^2 \sin^2 \left(
{ m \xi_2 \s \over 2 \hbar t} \right)
&(2.28) \cr }
$$
In the limit of small $\s$, the last two factors, arising from the
integration over $X_1$ and $X_3$, are approximately equal to $\s^2$.
Reverting to the variables $x = X_2 + \half \xi_2$, $y  = X_2 -
\half \xi_2$, Eq.(2.28) may then be written in the form,
$$
p(\F_2) = \left( { m \s \over \pi \hbar t} \right)^2
\Biggl| \int_{-{\s \over 2}}^{{\s \over 2}}
dx \ \exp \left( {2 i m \over \hbar t} x^2
- {it \over \hbar} \F_2 x \right) \Biggr|^2
\eqno(2.29)
$$
Now as discussed earlier, small $\s$ means
$ \s^2 \ << \ \hbar t / m $. Since $|x| < \s/2 $ in the integrand
of (2.29), the term quadratic in $x$ will therefore be negligible.
We may then approximately evaluate the integral, with the result,
$$
p(\F_2) \ \approx \ {4 m^2 \s^2 \over \pi^2  t^4  \F_2^2}
\ \sin^2 \left( { \s t \over 2 \hbar} \F_2 \right)
\eqno(2.30)
$$
We thus obtain agreement with the case of approximate
projections: the distribution of $\F_2$ is concentrated around
$\F_2=0$, with a width $\Delta F \approx { \hbar \over \s t}$
for small $\s$.

\head{\bf III. General Potentials.}

In the last section we exhibited the peaking about classical paths
for sequences of position samplings of the form (2.1), and
determined the width of the peaking. That analysis was for the free
particle. In this section, we extend it to the case of a particle in
a general potential.

Our starting point is again the expression (2.9) for a sequence of
position samplings. For general potentials, the density matrix
propagator $J$ cannot be given in closed form. We will therefore use
the WKB approximation for the propagator,
$$
\la x^{\pp}, t^{\pp} | x', t' \ra \approx  \Delta( x^{\pp}, t^{\pp} | x',
t') \ \exp \left( \ih S ( x^{\pp}, t^{\pp} | x', t' ) \right)
\eqno(3.1)
$$
Here, $S$ is the classical action between initial and final points.
It satisfies the Hamilton-Jacobi equation,
$$
{ 1 \over 2m} \left( { \partial S \over \partial x^{\pp} }
\right)^2 + V(\xpp) = - { \partial S \over \partial t^\pp}
\eqno(3.2)
$$
and similarly for the initial point $x',t'$.
The prefactor $\Delta$ is given by,
$$
\Delta (x^{\pp}, t^{\pp} | x', t')
= \left[ - {1 \over 2 \pi i \hbar} { \partial^2
S(x^{\pp}, t^{\pp} | x', t') \over \partial x^{\pp} \partial x'}
\right]^{\half}
\eqno(3.3)
$$
The approximation (3.1) will be exact for linear systems, or in the short
time limit.

The Hamilton-Jacobi equation may be solved in the short time limit
for general potentials. The solution, which will be needed later, is
$$
S(\xpp, t | x', 0 ) = {A(\xpp,x') \over t} + B(\xpp,x') \ t + O(t^3)
\eqno(3.4)
$$
where
$$
\eqalignno{
A(\xpp,x') &= \half m (\xpp -x')^2
&(3.5) \cr
B(\xpp,x') &= - { 1 \over ( \xpp - x') }
\int_{x'}^{\xpp} dx \ V(x)
&(3.6) \cr }
$$

We will use Gaussian projections. As we have seen, the case of
exact projections is not qualitatively different.
In the WKB approximation (3.1),
the probability for a sequence of position sampling is
$$
\eqalignno{
p(\x_1, \x_2, \cdots \x_n) & \approx \int \ \prod_{k=1}^n dx_k dy_k
\ \delta(x_n -y_n) \ \rho (x_1, y_1)
\cr & \times \exp \left( \ih \sum_{k=1}^{n-1} \left[
S(x_{k+1}, t_{k+1} | x_k, t_k) -
S(y_{k+1}, t_{k+1} | y_k, t_k) \right] \right)
\cr & \times
\exp \left( -
\sum_{k=1}^n { (x_k -\x_k)^2 \over 2 \s_k^2 }
- \sum_{k=1}^n { (y_k -\x_k)^2 \over 2 \s_k^2 }
\right)
&(3.7) \cr }
$$
(we do not include the prefactors explicitly).
This expression becomes exact in the limit that the time separation
between projections goes to zero, and (3.7) then becomes a path integral,
$$
\eqalignno{
p[\x(t)] &= \int {\cal D} x {\cal D} y \ \delta (x_f -y_f)
\ \rho(x_0, y_0)
\cr & \times \exp \left( \ih S[x(t)] - \ih S[y(t)] \right)
\cr & \times \exp \left(
- \int dt \ { (x(t) - \x(t) )^2 \over 2 {\tilde \s}^2 }
- \int dt \ { (y(t) - \x(t) )^2 \over 2 {\tilde \s}^2 }
\right)
&(3.8) \cr }
$$
(we use $x_0$, $x_f$, \etc, to denote initial and final values in the path
integral expressions).
Here, as in the previous section, $\tilde \s(t) $ is the limit of
$ t^{\half} \s_k $, as $t \ria 0$, and $\s_k \ria \infty$.
In what follows it will be convenient to work with both of these
expressions. As in the case of the free particle, we shall exhibit
the peak about the classical fields equations, compute the width of
the peak, and find the limits on it imposed by the uncertainty
principle.

Eqs.(3.7), (3.8) cannot be evaluated exactly, in general, except for
the case of linear systems, where all the integrals are Gaussians.
However, the dominant behaviour may be extracted in the limits of
large and small $\ts$.

For large $\ts$, the action terms in (3.8) are allowed to oscillate
rapidly, and by the stationary phase approximation, the paths
dominating the path integral will be the classical paths connecting
the initial and final points. Eq.(3.8) will therefore be
approximately equal to
$$
\eqalignno{
p[\x(t)] &= \int  dx_f dy_f dx_0 dy_0 \ \delta (x_f -y_f)
\ \rho(x_0, y_0)
\cr & \times \exp \left( \ih S(x_f, T | x_0, 0)  - \ih S(y_f, T |
y_0, 0 ) \right)
\cr & \times \exp \left(
- \int dt \ { (x_{cl}(t) - \x(t) )^2 \over 2 {\tilde \s}^2 }
- \int dt \ { (y_{cl}(t) - \x(t) )^2 \over 2 {\tilde \s}^2 }
\right)
&(3.9) \cr }
$$
Here $T$ is the total time duration of the history, $x_{cl}(t)$
denotes the classical path with $x_{cl}(0) = x_0$, $x_{cl}(T) =
x_f$, and similarly for $y_{cl}(t)$. It is difficult to evaluate this
expression further, except for linear systems. However, it is clear
that is peaked when $\x(t)$ lies along a classical path.
Furthermore, the width of the peak about the classical path will be
due almost entirely to the imprecision $\ts$ in the specification of
the path $\x$, and not due to quantum uncertainty. Given this, the
width $\Delta F$ of the peak about $F=0$ may in principle be
estimated, but we shall not do so here.

Now consider the case of small $\s$, and consider Eq.(3.7). As
before, introduce the variables $X= \half (x+y)$, $\xi = x-y$.
Eq.(3.7) may therefore be written,
$$
\eqalignno{
p[\x]  &= \int \ \prod_{k=1}^n dX_k d\xi_k
\ \delta(\xi_n) \ \rho (X_1+ \half \xi_1, X_1- \half \xi_1)
\cr & \times \exp \left( \ih \sum_{k=1}^{n-1} \left[
\xi_{k+1} { \partial S \over \partial X_{k+1} }
(k+1|k) +
\xi_k {\partial S \over \partial X_k}
(k+1|k) + O(\xi^3) \right] \right)
\cr & \times
\exp \left( -
\sum_{k=1}^n { (X_k -\x_k)^2 \over \s_k^2 }
- \sum_{k=1}^n { \xi_k^2 \over 4 \s_k^2 }
\right)
&(3.10) \cr }
$$
where we have introduced the notation,
$S(k+1|k) =  S(X_{k+1}, t_{k+1} | X_k, t_k) $.
Here, we have expanded the action terms about $\xi_k=0$, using the
fact that small $\s_k$ in the Gaussian projection
concentrates the contribution from $\xi_k$ to
the neighbourhood of $\xi_k = 0$.
Next we introduce the Wigner function of the initial density matrix,
via (2.11), and rearrange the actions terms using the discrete
analogue of an integration by parts:
$$
\eqalignno{
p[\x] &= \int \ \prod_{k=1}^n dX_k d\xi_k dp_1
\ \delta(\xi_n) \ W(p_1, X_1)
\ \exp \left( \ih \xi_1 \left[ p_1 + { \partial S \over \partial
X_1} (2|1) \right] \right)
\cr & \times \exp \left( \ih \sum_{k=2}^{n-1} \xi_k \left[
{ \partial S \over \partial X_{k} } (k|k-1)
+ {\partial S \over \partial X_k} (k+1|k) \right]
\right)
\cr & \times
\exp \left( - \sum_{k=1}^n { (X_k -\x_k)^2 \over \s_k^2 }
- \sum_{k=1}^n { \xi_k^2 \over 4 \s_k^2 } \right)
&(3.11) \cr }
$$
The integration over $\xi_k$ may now be performed, and one obtains,
$$
\eqalignno{
p[\x] &= \int \ \prod_{k=1}^n dX_k dp_1
\ \ W(p_1, X_1)  \ \exp \left( - {\s_1^2 \over \hbar^2}
\left[ p_1 + { \partial S \over \partial
X_1} (2|1) \right]^2 \right)
\cr & \times \exp \left( - \sum_{k=2}^{n-1} { \s_k^2 \over \hbar^2}
\left[
{ \partial S \over \partial X_{k} } (k|k-1)
+ {\partial S \over \partial X_k} (k+1|k) \right]^2
\right)
\cr & \times
\exp \left( - \sum_{k=1}^n { (X_k -\x_k)^2 \over \s_k^2 } \right)
&(3.12) \cr }
$$

We would like now to take the continuum limit of this expression.
Let $t_{k+1}-t_k =t$, and let $\ts^2_k = t \s_k^2$, as in Section 2.
Then we let $t \ria 0$ and $\s_k \ria \infty$ in such a way that
$\ts_k$ remains finite. Now consider the derivatives of the
Hamilton-Jacobi function appearing in Eq.(3.12). From
Eqs.(3.4)--(3.6), it may be shown that in the limit $t \ria 0$,
$$
\eqalignno{
- { \partial S \over \partial X_1} (2|1) \ & \ria \  m \dot X_0
&(3.13) \cr
- { 1 \over t} \left( { \partial S \over \partial X_{k} } (k|k-1)
+ {\partial S \over \partial X_k} (k+1|k) \right) \ & \ria
\ m \ddot X + V'(X)
&(3.14) \cr }
$$
Now taking the limit, one obtains,
$$
\eqalignno{
p[\x(t)]  = \int {\cal D} X \ & W( m \dot X_0, X_0) \ \exp \left( - \int
dt \ { ( X- \x )^2 \over \ts^2 }\right)
\cr & \times
\exp \left( - \int dt \ { \ts^2 \over \hbar^2} \ \left( m \ddot X +
V'(X) \right)^2 \right)
&(3.15) \cr }
$$
This is the probability for a history characterized by continuous
position samplings. This expression is exact for linear systems.

It might seem that there is a contradiction between the continuum
limit used above, which involves $\s_k \ria \infty$, and the
approximation, for non-linear systems,
used in deriving (3.10), which assumed small $\s_k$.
The point is, however, that the derivation of (3.10) requires that
the Gaussian in $\xi_k$ be strongly concentrated about $\xi_k=0$. In
the continuum limit, this will be assured if $\ts$ is small.
For non-linear systems, Eq.(3.15) is therefore valid for $\ts$
sufficiently small to guarantee the validity of the expansion about
$\xi_k=0$ used to derive Eq.(3.10).

The sum in (3.15) is over all paths $X(t)$ and the
integral is weighted by the Wigner function of the initial data of
the paths, $X_0$, $m \dot X_0$. It strongly suggests that $p[\x]$
will be peaked about paths satisfying the field equations, and that
the width $\Delta F$ of the peak is greater than about $\hbar / \ts $.
These
assertions are not completely watertight, however,
partly because the
Wigner function is not positive in general.
To make them so,
we need to compute the distribution of the value of the field
equations, $\F$, as in Section 2.

To do this, we go back to Eq.(3.11) and take the continuum limit.
The result is,
$$
\eqalignno{
p[\x] &= \int {\cal D}X {\cal D} \xi dp_0 \ \delta (\xi_f) \ W(p_0, X_0)
\ \exp \left( \ih p_0 x_0 \right)
\cr & \times
\exp \left( - \ih \int dt \ \xi \left( m \ddot X + V'(X) \right) -
\ih \xi_0 M \dot X_0 \right)
\cr & \times
\exp \left( - \int dt \ { (X- \x)^2 \over \ts^2 } - \int dt \ { \xi^2
\over 4 \ts^2} \right)
&(3.16)\cr}
$$
Next we let $X \ria X+ \x $, $ p_0 \ria p_0 + m \dot \x_0 $, and
then perform an integration by parts of the $\ddot X$ term,
with the result,
$$
\eqalignno{
p[\x] &= \int {\cal D}X {\cal D} \xi dp_0 \ \delta (\xi_f)
\ W(p_0 + m \dot \x_0, X_0 + \x_0)
\ \exp \left( \ih p_0 x_0 \right)
\cr & \times
\exp \left( \ih  m {\dot \xi}_f X_f - \ih m {\dot \xi}_0 X_0 \right)
\cr & \times
\exp \left( - \ih \int dt \ \xi \left( m \ddot \x + V'(\x) \right)
-\ih \int dt \left( m \ddot \xi X + \xi V'(X+\x ) - \xi V'(\x) \right) \right)
\cr & \times
\exp \left( - \int dt \ {X^2   \over \ts^2 } - \int dt \ { \xi^2
\over 4 \ts^2} \right)
&(3.17) \cr}
$$
The term $V'(X+\x)$ could be a source of some difficulty. However, we
again use the fact that $\ts$ is small, and hence the Gaussian
projector concentrates the contribution from the integral over $X$ to
the region around $X=0$. We may therefore make the approximation,
$$
V'(X+\x) \approx V'(\x) + X V^{\pp}(\x) + O(X^2)
\eqno(3.18)
$$
Now we may carry out the integral over $X$. The integrals over the
initial and final values pull down delta-functions, and we regard
the action terms as residing on the internal slices only
(as described in
more detail using the explicit time slicing in Section 2). One
obtains,
$$
\eqalignno{
p[\x] &= \int {\cal D} \xi dp_0 \ \delta (\dot \xi_f) \delta (\dot
\xi_0) \delta( \xi_f) \ W(p_0 + m \dot \x_0, X_0 + \x_0) \ \exp
\left( \ih p_0 \xi_0 \right)
\cr & \times
\exp \left( -ih \int dt \ \xi \F - \int dt \ { \xi^2
\over 4 \ts^2 } \right)
\cr & \times
\exp \left( - \int dt \ { \ts^2 \over 4 \hbar^2} \left( m \ddot \xi
+ \xi V^{\pp} (\x) \right)^2  \right)
&(3.19) \cr }
$$
where $ \F = m \ddot \x + V'(\x) $.

Now we would like to integrate out $\x_0$ and $m\dot \x_0$ in
$p[\x]$ to obtain $p[\F]$, the distribution on the field equations.
To do this, we are obliged to assume that $V^{\pp}(\x)$ has limited
dependence on $\x$. This will be good for systems close to linear
systems. Also, the dependence on $V^{\pp}(\x)$ will not matter in
the limit of small $\ts$. We thus obtain,
$$
\eqalignno{
p[\F] &= \int d\x_0 d(m\dot \x_0) \ p[\x]
\cr & =  \int {\cal D} \xi \ \delta (\dot \xi_f) \delta (\dot \xi_0)
\delta( \xi_f) \delta ( \xi_0)
\ \exp \left( -\ih \int dt \ \xi \F \right)
\cr \times \exp & \left(
- \int dt \ \xi \left(
{ m^2 \ts^2 \over 4 \hbar^2} { d^4 \over dt^4}
+ { m \ts^2 V^{\pp} \over 2 \hbar^2} {d^2 \over dt^2}
+ {\ts^2 {V^{\pp}}^2 \over 4 \hbar^2} + { 1 \over 4 \ts^2}
\right) \xi \right)
&(3.20) \cr}
$$
where an integration by parts has been performed, and the boundary
conditions invoked to drop the boundary terms.
Finally, the integral over $\xi$ may be performed to yield
a result identical in form to Eq.(2.21), but now $G(t,t')$ is the
Green function of the fourth order operator appearing in (3.20).
One may thus in principle compute the width of the peak. We shall
not carry this out explicitly, but is clear that the width will
again satisfy,
$$
(\Delta \F)^2 \ge { \hbar ^2 \over 2 \ts^2 }
\eqno(3.21)
$$

We could in principle have deduced these results from Eq.(3.15)
directly. However, we chose to do it this way, taking Eq.(3.11) as
our starting point, because the boundary conditions in the resulting
path integral expression (3.20), are much simpler than those on
$X(t)$ in (3.15).

Mensky has previously given a similar result on the basis of a
rather heuristic analysis [\cite{mensky1}]. He argued that the
uncertainty principle for histories consisting of continuous
position samplings takes the general form,
$$
\Delta({\rm path}) \ \Delta ( {\rm field \ equation} ) \ \ge \ \hbar
\eqno(3.22)
$$
where $\Delta ( {\rm path}) $ is the width of the specification of
the path (here denoted $\s$), and $\Delta ({\rm field \ equation})$
is the spead about the classial field equations (here denoted
$\Delta F$).
We thus confirm his heuristic analysis.

In the case of the free particle, we were able to obtain an
expression for the width of the peak about $\F=0$ that was
completely independent of the initial conditions (for Gaussian slits).
Here the
dependence on $\x$ through $V^{\pp}(\x)$ in Eq.(3.19) meant that
it was not possible to integrate out the initial data as we did in
Section 2, and obtain complete independence of the initial
state. Unlike the free particle case, therefore, the presence of a
potential means that some dependence on the initial conditions
appears in the expression for the width of the peak about $\F=0$
(although it drops out in the limit of very small $\ts$).

\head {\bf IV. Systems with Thermal Fluctuations}

We now consider a straighforward extension of the result of the
previous sections to include thermal fluctuations. Suppose our
system is linearly coupled to an environment, consisting of a bath of
harmonic oscillators in a thermal state. Such models are frequently
studied in the context of decoherence studies, and non-equilibrium
statistical mechanics generally. We will give only the briefest of
details here, since these models have been discussed extensively
elsewhere
[\cite{gell2,brun,caldeira,dekker,grabert,hupaz,vernon,unruh}].

In the presence of thermal fluctuations, one would still expect the
probabilities for sequences of position samplings to become peaked
about classical paths, but the width of the peak will be broadened.
Our aim is to show this explicitly, and determine the comparative sizes
of quantum and thermal fluctuations.

It is straighforward to show that in the presence of an environment,
the probabilities for histories are given by
$$
\eqalignno{
p[\x(t)] &= \int {\cal D} x {\cal D} y \ \delta (x_f -y_f)
\ \rho(x_0, y_0)
\cr & \times \exp \left( \ih S[x(t)] - \ih S[y(t)] + \ih W[x,y] \right)
\cr & \times \exp \left(
- \int dt \ { (x(t) - \x(t) )^2 \over 2 {\tilde \s}^2 }
- \int dt \ { (y(t) - \x(t) )^2 \over 2 {\tilde \s}^2 }
\right)
&(4.1) \cr}
$$
This differs from Eq.(3.8) by the presence of the Feynman-Vernon
influence functional phase, $W[x,y]$, given by [\cite{vernon}],
$$
\eqalignno{
W[x(t),y(t)] = & -
\int_0^t ds \int_0^s ds' [ x(s) - y(s) ] \ \eta (s-s') \ [ x(s') + y(s') ]
\cr &
+ i \int_0^t ds \int_0^s ds' [ x(s) - y(s) ] \ \nu(s-s') \ [ x(s') - y(s') ]
&(4.2) \cr }
$$
The explicit forms of the non-local kernels $\eta$ and $\nu$ may be found
in Refs.[\cite{caldeira,hupaz}]. We have assumed, as is typical in
these models, that the initial density matrix of the total system is
simply a product of the initial system and environment density
matrices,
and the initial environment density matrix is a thermal state at
temperature $T$.
Considerable simplifications occur in a purely ohmic environment in
the Fokker-Planck limit (a particular form of the high temperature limit),
in which one has
$$
\eqalignno{
\eta(s-s') &=  m\gamma \ \delta '(s-s')
&(4.3) \cr
\nu(s-s') &= { 2 m \gamma k T \over \hbar } \ \delta (s-s')
&(4.4) \cr }
$$
where $\gamma$ is the dissipation.

The presence of the influence functional phase does not
qualitatively change the derivation given in Section 3, and
repeating the derivation, it is straighforward to show that
one has, in place of (3.19), (3.20),
$$
\eqalignno{
p[\F] & =  \int {\cal D} \xi \ \delta (\dot \xi_f) \delta (\dot \xi_0)
\delta( \xi_f) \delta ( \xi_0)
\cr & \times
\ \exp \left( -\ih \int dt \ \xi \F - \int dt dt' \ \xi(t) \left(
{ 1 \over 4 \ts^2} \delta (t-t') + { 1 \over \hbar} \nu(t-t')
\right) \xi(t')  \right)
\cr & \times
\exp \left( - \int dt  \ { \ts^2 \over 4 \hbar^2}
\left( m \ddot\xi(t) + V^{\pp}(\x) \xi(t) - \int dt' \eta(t-t')
\xi(t') \right)^2 \right)
&(4.5) \cr }
$$
where
$$
\F = m \ddot\x(t) + V'(\x) - \int dt' \eta(t-t') \x(t')
\eqno(4.6)
$$
A new feature of the derivation of (4.5) is that the presence of the
noise kernel $\nu(t-t')$ considerably enhances the Gaussian peak about
$\xi=0$ in (3.10), and thus the expansion about $\xi=0$ has a
greater range of validity, namely, it is no longer restricted
to the small $\ts$ regime. The validity of the expansion
about $X=0$ in (3.18), however, is not improved.

The integral over $\xi$ is readily carried out to yield a
result of the form (2.21).
The distribution $p[\F]$ will therefore be peaked about the
classical field equations, $\F = 0$,
but modified by an extra term, in Eq.(4.6), which
induces dissipation, and also renormalizes the potential.
Furthermore, the width of the peak is changed. In the limit of
either small $\ts$, or that in which $\nu(t)$ is large ({\it e.g.},
high temperature), or both, one has
$$
\la \F(t) \F(t') \ra = { \hbar^2 \over 2 \ts^2} \ \delta (t-t')
+ 2 \hbar \nu(t-t')
\eqno(4.7)
$$
where the left-hand side denotes an average in the distribution
function (4.5). In the Fokker-Planck limit, one has
$$
\la \F(t) \F(t') \ra \ \approx \ 4 m \gamma kT \ \delta (t-t')
\eqno(4.8)
$$

These results are consistent with regarding the system, in this limit,
as described by a classical Langevin equation of the form,
$$
m \ddot \x + m \gamma \dot \x + V_R^{\prime}(\x) = n(t)
\eqno(4.9)
$$
where $V_R$ is the renormalized potential, and $n(t)$ is a
fluctuating force
term with $ \la n(t) \ra = 0 $ and two-point correlation function
given by (4.8). We therefore recover classical Brownian motion in the
Fokker-Planck limit.

Similar results have been derived using different methods by
Gell-Mann and Hartle [\cite{gell2}].
They used a distribution $g[\x(t), \F(t)]$
on $\x(t)$ and $\F(t)$, obtained by a type of Wigner transform of
the decoherence functional. Like the Wigner function, their
distribution is not always positive. Here, by contrast, we have used
an explicitly positive distribution on the value of the field
equations, $p[\F(t)]$. Also, the derivation given here exposes the
necessity, not evident from Ref.[\cite{gell2}],
to make the approximation (3.18) in the case of
non-trivial potentials.

It is perhaps of interest to remark that the existence of the distribution
$g[\x(t),\F(t)]$ used by Gell-Mann and Hartle
suggests, in analogy with the ordinary Wigner
function $W(p,q)$, that $\x(t)$ and $\F(t)$ are some kind of
``canonically conjugate pair''. If this could be made precise, it
might supply the underlying reason why
the uncertainty principle for continuous histories
has the form (3.21), involving a product of the widths of $\x$ and $\F$.
These possibilities will be investigated elsewhere.

For us, the significance of the result (4.7) is that it
indicates the respective regimes in which quantum or thermal
fluctuations dominate. In the Fokker-Planck limit, the thermal
fluctuations dominate the quantum ones when
$$
{ 8 m \gamma k T \ts^2 \over \hbar^2 } \ >> \ 1
\eqno(4.10)
$$
This is in fact also the condition for decoherence
in quantum Brownian motion models
[\cite{gell2,dowker,paz,zurekpazhabib,zurek2}]. We therefore find
that the regime in which thermal fluctuations dominate the quantum
ones coincides with the onset of decoherence. Essentially the same
conclusion was reached in  Ref.[\cite{ander}], using a quite
different measure of uncertainty (the Shannon information used in
Eq.(5.8) below). This in turn built on the earlier work of Hu and
Zhang [\cite{huzhang}].

\head{\bf V. Alternative Characterizations of Classical Peaking}

We have so far characterized peaking about classical paths using a
distribution function on the value of the field equations, Eq.(2.7).
Now we discuss an alternative method.
Consider the probability for a sequence of position samplings,
(2.1). As we have discussed before, let us replace $\x_1, \x_2$,
with $\x_1, \k_1 = m (\x_2 -\x_1)/t $. For sufficiently small $t$,
$\k_1$ is an approximation to the initial momentum.
Now consider the probability, $p(\x_1, \k_1, \x_3, \cdots \x_n)$,
defined using this change of variables. From it, we may
construct the conditional probability,
$$
p(\x_3, \cdots \x_n| \x_1, \k_1) = { p(\x_1, \k_1, \x_3, \cdots \x_n) \over
p(\x_1, \k_1) }
\eqno(5.1)
$$
This is the probability of finding the sequence $\x_3, \cdots \x_n$,
{\it given} that $\x_1, \k_1$ have already occurred. This
conditional probability distribution provides an alternative measure
of classical peaking: the system exhibits
classical correlations if (5.1) is strongly peaked about the
configuration for which $\x_3 \cdots \x_n$ lie along the classical
path with initial data $\x_1, \k_1$. We now show how this is related
to the condition that Eq.(2.7) is peaked about $\F=0$.

For simplicity, we use samplings continuous in time,
although we expect that
the results and approach are more general than this.
Consider the probability for a history of continuous position
samplings, $p[\x(t)]$. Instead of $p[\x(t)]$, consider the distribution
$p[\F(t);\x_0, \k_0)$ on the field equations $\F = m \ddot \x +
V'(\x)$ together with the initial data, $\x_0, \k_0$. This
distribution is a functional of $\F(t)$, and a function of $\x_0,
\k_0$, and we use the notation $[\ ; \ )$ to indicate this.
It may be given explicitly in terms of $p[\x(t)]$ by the
functional integral expression,
$$
p[\F(t);\x_0, \k_0) = \int {\cal D} \x \ p[\x(t)]
\ \delta(\x(0) -\x_0)
\ \delta (m \dot \x(0) - \k_0 ) \ \delta \left[ m \ddot \x + V'(\x)
- \F \right]
\eqno(5.2)
$$
Introducing the continuum analogue of (5.1), we may write,
$$
p[\x(t)] = p[\x(t) | \x_0, \k_0 ) \ p(\x_0, \k_0)
\eqno(5.3)
$$
so $p[\x(t)|\x_0,\k_0)$ is the probability of finding the history
$\x(t)$, given that its initial data are $\x_0, \k_0$.
Inserting this in (5.2), we have for the distribution on the field
equations,
$$
\eqalignno{
p[\F(t)]  \equiv & \int d \x_0 d \k_0 \ p [\F(t); \x_0, \k_0 )
\cr  =&
\int  {\cal D} \x \ d \x_0 d \k_0 \ p[\x(t)| \x_0, \k_0 ) \ p(\x_0, \k_0)
\cr & \times
\ \delta ( \x(0) - \x_0 ) \ \delta (m \dot \x(0) - \k_0 )
\ \delta \left[ m \ddot \x + V'(\x) - \F \right]
&(5.4) \cr }
$$
Now we have shown extensively in previous sections that $p[\F(t)]$
is strongly peaked about $\F(t) = 0$. In other words, $p[\F(t)]$ is
essentially zero unless, $\F$ is very close to zero. Using (5.4),
this means that
$$
\eqalignno{
\int & {\cal D} \x \ d \x_0 d \k_0 \ p[\x(t)| \x_0, \k_0 ) \ p(\x_0, \k_0)
\cr & \times
\ \delta ( \x(0) - \x_0 ) \ \delta (m \dot \x(0) - \k_0 )
\ \delta \left[ m \ddot \x + V'(\x) - \F \right] \ \approx \ 0
&(5.5) \cr }
$$
unless $\F$ is very close to zero. Now the terms in the integrand
are all positive, and generally non-zero with the exception of
$p[\x(t)| \x_0, \k_0)$. This means that
$$
p[\x(t)| \x_0, \k_0) \ \approx \ 0
\eqno(5.6)
$$
unless $\F = 0$, \ie, unless $ \x(t) $ satisfies the field
equations. Furthermore, the delta-functions ensure that
$\x(0) = \x_0$ and $m\dot \x(0) = \k_0 $. We therefore conclude that
if $p[\F(t]$ is peaked around $ \F(t) = 0$, then
$p[\x(t)| \x_0, \k_0)$ is peaked about the solution $\x(t)$ to the
field equations with initial data $\x_0, \k_0$.

It is not hard to see that the converse is also true. If $p[\x(t)|
\x_0, \k_0) $ is strongly peaked about the classical
solution $\x(t)$ with
initial data $\x_0, \k_0$, then the functional integral in (5.4)
receives contributions only from paths $\x(t)$ such that
$ m \ddot \x + V'(\x) \approx  0 $. From the delta-functional
in (5.4) it follows that $p[\F(t)]$
will be concentrated around $\F(t)=0$.

We now briefly comment on the formal interpretation of the
probabilities for strings of position samplings considered in
previous sections. It is tempting to interpret expressions such as
(3.15) as the statement that the system is described by a
statistical ensemble of classical solutions,
with the probability for each individual solution given by
the Wigner function of the initial density matrix.

This cannot be completely correct. Firstly, the Wigner function is
not always positive. Secondly,
the coarse-grained histories $\x(t)$ are usually taken to be a
decoherent set, \ie, they satisfy the probability sum rules.
Regarding the probability $p[\x]$ for each history $\x(t)$ as a sum
over initial data of probabilities for histories with each possible
value of initial data corresponds to a {\it fine graining} of the
histories, under which decoherence is not preserved. The above
statement therefore requires a more careful formulation.

Let us start with the conditional probability (5.1).
With some
elementary rearrangement, we may write,
$$
\eqalignno{
p(\x_3, \cdots \x_n ) & \equiv \int d\x_1 d\k_1\ p(\x_1, \k_1,
\x_3, \cdots \x_n )
\cr & = \int d \x_1 d \k_1
\ p(\x_3, \cdots \x_n| \x_1, \k_1) \ p(\x_1, \k_1)
&(5.7) \cr }
$$
Consider the two parts of the integrand in Eq.(5.7).  First of all
consider the conditional probability. As we have seen above
in the continuum case, this
is strongly peaked about configurations $\x_3, \cdots \x_n $ lying
along the classical path with initial data $\x_1, \k_1 $.

Next consider the quantity $p(\x_1, \k_1)$. It is positive, by
construction, and may be interpreted as a quantum mechanical
probability distribution for coarse-grained phase space samplings.
It is in fact equal to a {\it smeared} Wigner function, with a
smearing just sufficient to make it positive [\cite{jjh}].
Furthermore, the
uncertainty principle imposes limits on the degree to which it may
be peaked about a given region of phase space. This limit may be
expressed as a lower bound on its Shannon information,
$$
I(X,K) \equiv - \int d\x d\k \ p(\x, \k) \ln p(\x, \k) \ \ge \ \ln
\left( { 2 \pi \hbar \over \s_x \s_k } \right)
\eqno(5.8)
$$
where $\s_x$, $\s_k$ are the widths of the position and momentum
samplings [\cite{jjh2}]. More stringent limits on the degree of peaking arise
in the presence of thermal fluctuations (as discussed in the
previous section); these may also be expressed as a lower bound on
the Shannon information, generally greater than the lower bound in
(5.8) [\cite{ander}].

Now we may say the following
above about Eq.(5.7): it may be interpreted as the statement that
$p(\x_3, \cdots \x_n)$ corresponds to a statistical ensemble of
classical solutions, where the probability for each individual
solution is given by the smeared Wigner function, $p(\x_1, \k_1)$.
The original statement is therefore in essence correct, but its
precise formulation requires coarse-graining over the first two
position samplings.

\head{\bf VI. Decoherence and Measurements}

So far, we have largely been concerned in this paper with the
mathematical properties of the expression (1.1).
As stated in the
Introduction, the results of Section 2--5 are not necessarily tied
to a particular interpretational scheme, be it the decoherent
histories approach, or the Copenhagen approach to measured
subsystems.
We now discuss these points in relation to the considerations of
this paper. In particular, we need to discuss the extent to which
the histories discussed in Sections 2--5 satisfy the consistency
condition, that the decoherence functional (1.5) be diagonal:
$$
D( {\au} , {\au'} ) \approx 0, \quad {\rm for} \quad {\au \ne \au'}
\eqno(6.1)
$$

\subhead{\bf 6(A). The Significance of the Uncertainty Relations for
Histories }

Consider a closed system and consider possible histories of
alternatives of that system characterized by strings of projections
operators. A given set of histories will generally not satisfy the
decoherence condition (6.1), and probabilities cannot be assigned to
the histories. As is well known, decoherence is generally brought
about by dividing the total closed system into a distinguished
subsystem coupled to the rest -- the
environment -- and considering histories
characterized by projections onto the distinguished subsystem only.
An example of this type of model was discussed in Section 4.  As
stated there, decoherence in this model is achieved when the
parameters satisfy Eq.(4.10). Given a decoherent set of histories,
one can then begin to discuss classical predictability, and this
indeed we did in Section 4. However, the usual quantum fluctuations
about classical predictability are accompanied
by thermal ones, due to
the coupling to the environment. As we saw in Section 4, the thermal
fluctuations typically dominate in the regime in which there is
decoherence. This means that the uncertainty principle plays little
{\it physical} role in situations of this type, since the quantum
fluctuations are completely swamped by the thermal ones.

As the coupling to the environment goes to zero, the thermal
fluctuations go to zero leaving just the quantum ones, but the
degree of decoherence goes to zero also. The probabilities (1.1)
then no longer satisfy the decoherence condition (6.1). However, the
uncertainty relations such as (3.21) still maintain some
mathematical utility, in that they are relations that must be
satisfied by the candidate  probabilities (1.1) in the limit that
the coupling to the environment goes to zero. The fact that the
probabilities are not decoherent does not matter mathematically,
because at no stage were the probability sum rules assumed
in deriving (3.21).

In the decoherent histories approach, the projections are not
associated with measurements by an external agency. They cannot be
because the system is genuinely closed. Rather, they are the way in
which histories of the system are specified. The uncertainty
relations, with or without thermal modifications, represent
fundamental limitations on the precision with which the properties
of system are {\it intrinsically defined}\footnote{$^{\dag}$}{See
Ref.[\cite{hartle2}] for more discussion of this and related
points.}.

There is an alternative way of thinking about the
probabilities (1.1) in which the uncertainty relations such as
(3.21), without thermal fluctuations, play a more significant role.
This is the special case of a measurement situation.
Again consider a closed quantum system, so we are still in the
framework of the decoherent histories approach.
Let the system consist of
of a distinguished subsystem
$\S$, a measuring apparatus $\A$, and the rest, the environment,
$\E$. The subsystem $\S$ is isolated from the environment, but
interacts in a very particular way with the apparatus
$\A$ and thus becomes
correlated with it. The apparatus, which is typically macroscopic,
interacts with the environment. As a consequence,
histories of apparatus alternatives decohere.  Histories of the
distinguished subsystem alternatives then also decohere, because
they are correlated with decohering apparatus alternatives.
Furthermore, under these conditions it may then be shown that the
probabilities for the sequence of decohering subsystem alternatives
is given by (1.1) [\cite{hartle,griffiths,omnes}].

We can therefore think of the probabilities of the form (1.1) as the
probabilities of a decoherent set of histories, if we regard them as
the probabilities for a sequence of {\it measured} alternatives of a
distinguished subsystem. Indeed, the formula (1.1) is a familiar
formula in the Copenhagen approach to measured subsystems: it
incorporates both the ``collapse of the wave function'', when a
measurement is made, together with the unitary evolution between
measurements [\cite{wigner}]. Here, it emerges from the decoherent
histories approach in the idealization of perfect correlation with
the apparatus and perfect decoherence of the apparatus alternatives
[\cite{hartle}].

For us, the point of this is that the interaction between the
distinguished subsystem and the measuring apparatus is not an
arbitrary one, but is
in principle
carefully designed so as to eliminate the environmental
fluctuations suffered by the system described in Section 4.
Measurements could therefore be contemplated in which
the lower limit (3.21) is approached. The uncertainty
relations such as (3.21) then do have physical significance; indeed,
they have their
more familiar significance as
limitations on the precision with which certain quantities may be
actually measured by an apparatus.

\subhead {\bf 6(B). The Decoherence Functional}

We now study in more detail the question of decoherence of the histories
considered in Section 2--5.
We are generally concerned with histories characterized by sequences
of position samplings, for which the decoherence functional is given
by
$$
D(\x_1, \cdots \x_n | \x_1', \cdots \x_n') = \Tr \left(
P_{\x_n}(t_n) \cdots P_{\x_1}(t_1) \ \rho \ P_{\x_1'}(t_1) \cdots
P_{\x_n'}(t_n) \right)
\eqno(6.2)
$$
It is convenient to introduce the notation,
$$
C_{\x_1 \cdots \x_n} = P_{\x_n}(t_n) \cdots P_{\x_1}(t_1)
\eqno(6.3)
$$
This object will be called a class operator. Histories characterized
by sequences of alternatives of the value of $\F_i$ are obtained by
coarse-graining the class operator (6.3). Precisely,
$$
C_{\F_2 \cdots \F_{n-1} } = \int d \x_1 \cdots d \x_n \
\prod_{k=2}^{n-1} \ \delta \left( m {(\x_{k+1}-2\x_k + \x_{k-1})
\over t^2} + V'(\x_k) - \F_k \right) \ C_{\x_1 \cdots \x_n}
\eqno(6.4)
$$
(where we are for convenience adopting a notation in which the
sampled quantities $\x$, $\F$, \etc, are continuous, but this is not
an essential assumption). This is, incidently,
an example of spacetime coarse-graining [\cite{hartle3}] -- a
coarse-graining that cannot be expressed as a sum of projection
operators at a single moment of time.
The decoherence functional for the $\F_i$'s is then
$$
D(\F_2 \cdots \F_{n-1} | \F_2' \cdots \F_{n-1}' ) = \Tr \left(
C_{\F_2 \cdots \F_{n-1}} \ \rho \ C^{\dag}_{\F_2' \cdots \F_{n-1}'} \right)
\eqno(6.5)
$$
Probabilities for histories are generally obtained from the diagonal
components of the decoherence functional, $D(\F, \F)$ (where we use
the shorthand $\F$ to denote a string of $\F_i$'s). In the earlier
sections of this paper, we obtained the probability  distribution
$p(\F_2, \cdots \F_{n-1})$ by integrating  out the initial data in
the probabilities, \ie, incoherent summing. In effect, we carried
out the operation (6.4) for the probabilities, not for the class
operators. The result is generally not the same as $D(\F,\F)$ which is
obtained by coherent summing. However, if the decoherence functional
is diagonal in the first two position samplings (and hence in the
initial data $\x_1,\k_1$),  then $p(\F) = D(\F,\F)$.  We shall assume
this. It has the following interesting consequence.

The decoherence functional obeys a useful inequality bounding the
size of the off-diagonal terms by the corresponding diagonal ones
[\cite{dowker}]. For (6.5) it is,
$$
\Bigl| D(\F,\F') \Bigr|^2 \ \le \ D(\F,\F) \ D (\F', \F')
\eqno(6.6)
$$
Its intuitive content is that there can be no
inteference with a history with $D(\F,\F)=0$. In particular,
it implies that
consistency is automatically satisfied if the system has one
history with $D(\F,\F) =1 $, and $D(\F',\F')=0$
for all the other histories.

As we have shown in this paper, the distribution  $p(\F_2 \cdots
\F_{n-1})$ is generally strongly peaked about $\F_i=0$. This means
that there is in essence only one history with non-zero probability:
that with $\F_i \approx 0 $ at each moment of time. From the
inequality (6.6), it is therefore readily seen that the off-diagonal
terms of the decoherence functional will be strongly suppressed --
the decoherence functional is essentially zero unless $ \F_i \approx
\F_i' \approx 0 $ for $i=2,\cdots n-1 $. We therefore see that
decoherence of the first two position samplings, together with the
peak about $\F_i = 0$ of $p(\F)$ ensures decoherence
of the $\F_i$'s.

Similar remarks hold for the probabilities defined by Eq.(5.7).
Consider the decoherence functional (6.2), and suppose a change of
variables from $\x_2$ to the initial momentum $\k_1$ is carried out,
as discussed before. Now consider the decoherence functional
obtained by coarse-graining over the initial data, $\x_1, \k_1$. It
is in general given by a sum over both off and on diagonal terms in
the initial data. However, if as before, we assume decoherence in
the initial data, then it is given by,
$$
D(\x_3, \cdots \x_n | \x_3' \cdots \x_n' ) = \int d\x_1 d \k_1 \
\ D( \x_1, \k_1, \x_3, \cdots \x_n | \x_1, \k_1, \x_3' , \cdots
\x_n' )
\eqno(6.7)
$$
Now we again use the inequality from Ref.[\cite{dowker}], of which
(6.6) was an example:
$$
\Bigl| D(\x_3, \cdots \x_n | \x_3' \cdots \x_n' ) \Bigr|
\ \le \ \int d \x_1  d \k_1
\ \left[
p( \x_1, \k_1, \x_3, \cdots \x_n )
p( \x_1, \k_1, \x_3', \cdots \x_n' )
\right]^{\half}
\eqno(6.8)
$$
Using the conditional probability defined by (5.1), this may be
written
$$
\eqalignno{
\Bigl| D(\x_3, \cdots \x_n | \x_3' \cdots \x_n' ) \Bigr|
\ \le \ & \int d \x_1  d \k_1 \ p(\x_1, \k_1)
\cr  & \times
\ \left[
p(\x_3, \cdots \x_n | \x_1, \k_1)
p(\x_3', \cdots \x_n'| \x_1, \k_1 )
\right]^{\half}
&(6.9) \cr}
$$
As we showed in Section 5, the conditional probability
$p(\x_3, \cdots \x_n | \x_1, \k_1)$ is strongly peaked about the
classical path with initial data $\x_1, \k_1$. That is, it is
essentially zero unless $\x_3, \cdots \x_n$ lie along that path.
It follows that the right-hand side of (6.9) will be very small
unless $\x_i \approx \x_i'$, for $i=3, \cdots n$. We therefore find
that the probabilities  (5.7) are decoherent, assuming the
decoherence of the initial data, and given the classical
predictability discussed in Section 5.

It is perhaps of interest to note that it is essentially the {\it
correlation} of $\x_3, \cdots \x_n$ with $\x_1, \k_1$ that leads to
their decoherence. It is often the case that decoherence of a given
subsystem arises as a result of its correlation with another
subsystem. This correlation may arise through interaction, although
need not necessarily. The demonstration of the connection between
correlation with another system and decoherence proceeds in a manner
similar to that given above [\cite{jjh3}]. Here, the correlation is
not with another system, but is between successive values of
positions and comes about as a consequence of the quantum dynamics.
One could, for example, imagine that the first two samplings are
actual measurements, or at least, interaction with another system.
The decoherence of future histories of position samplings is then
assured because of their correlation with the decohered initial data.

The considerations of this section are closely related to the
detailed proofs given by Omn\`es of classical determinism in quantum
mechanics [\cite{omnes,omnes2}]. He showed that an initial
quantum state $\rho_0$
localized in phase space to a cell $C_0$, will evolve under unitary
evolution to a state $\rho_t$ localized to $C_t$, the classical
evolution of $C_0$. There will of course be errors, due for
example to wave packet spreading, and these can be estimated.
But one would expect them to be very small indeed
if the cells are large compared to $\hbar$, if the particle is
sufficiently massive, and if the period of evolution is not too long.
Furthermore, sets of histories characterized by an initial state and
by (approximate) phase space projectors of the type indicated above
will be approximately consistent, essentially because there is only one
history with non-zero probability, and the discussion after Eq.(6.6)
applies. It is essentially for these reasons that one might have
anticipated the results of this section.

\head{\bf VII. Summary and Conclusions}

We have studied quantum-mechanical histories characterized by
sequences of position samplings. We have shown, in two different
ways that such samplings are strongly correlated according to
classical laws: the probabilities of such histories are strongly
peaked about classical histories. The uncertainty principle arises
as a lower bound on the width of this peak, and we have computed the
form of this lower bound explicitly. It generally has the form (1.9)
for samplings at a discrete set of times, and (3.21) for samplings
continuous in time. In the presence of thermal fluctuations it is
broadened by the amount indicated in Eq.(4.7). The corrections due
to thermal fluctuations give an idea of the comparative size of
quantum and thermal fluctuations: the latter become important under
the same conditions that decoherence effects become important. The
numerical value of these lower bounds is generally very small indeed
for macroscopic systems.  In a sentence, our results are therefore
consistent with the statement that the deterministic evolution of
macroscopic systems emerges as an approximate feature of
quantum mechanics, with an exceedingly small
error.

\head{\bf Acknowledgements}

I am very grateful to numerous colleagues for useful conversations,
including  Arlen Anderson, Carl Caves, Chris Fuchs, Jim Hartle,
Bei-Lok Hu,  Chris Isham, Raymond Laflamme, Seth Lloyd and
Wojciech Zurek. I would like to thank Bernhard Meister for his
useful comments and for a
critical reading of the manuscript.
Part of this work was carried out
during a visit to the Los Alamos National Laboratory. I am grateful
to Wojciech Zurek and his group there, for hospitality during my
visit. This work was supported by a University Research Fellowship
from the Royal Society.

\references

\def\pr{{\sl Phys. Rev.\ }}

\def\prep{{\sl Phys. Rep.\ }}

\def\rmp{{\sl Rev. Mod. Phys.\ }}

\def\pl{{\sl Phys. Lett.\ }}
\def\annp{{\sl Ann. Phys. (N.Y.)\ }}

\def\jsp{{\sl J. Stat. Phys.\ }}

\refis{albrecht} A. Albrecht, {\sl Phys.Rev.} {\bf D46}, 5504 (1092).
%{\it Investigating Decoherence in a Simple System.}

\refis{blencowe} M. Blencowe, {\sl Ann.Phys.} {\bf 211}, 87 (1991).
%{\it The Consistent Histories
%Interpretation of Quantum Fields in Curved Spacetime.}

\refis{ander} A.Anderson and J.J.Halliwell, ``An
Information-Theoretic Measure of Uncertainty due to Quantum and
Thermal Fluctuations'', Imperial College Preprint 92-93/25 (1993),
gr-qc 9304025. To appear in {\sl Phys.Rev.D}

\refis{brun} T.Brun, \pr {\bf D47}, 3383 (1993).

\refis{caldeira} A.O.Caldeira and A.J.Leggett, {\sl Physica} {\bf
121A}, 587 (1983).

\refis{caves} C.Caves, \pr {\bf D33}, 1643 (1986).
% Quantum mechanics of measurements distributed in time.
% A path-integral formulation.

\refis{dekker} H.Dekker, \pr {\bf A16}, 2116 (1977).

\refis{dowker} H.F.Dowker and J.J.Halliwell, \pr {\bf D46}, 1580
(1992).

\refis{gell}
M.Gell-Mann and J.B.Hartle, in {\it Complexity, Entropy and the
Physics of Information. SFI Studies in the Sciences of Complexity,
Vol. VIII}, edited by W.Zurek (Addison Wesley, Reading, MA, 1990).

\refis{gell2} M.Gell-Mann and J.B.Hartle, \pr {\bf D47}, 3345 (1993).

\refis{hartle}
J.B.Hartle, in {\it Quantum Cosmology and Baby Universes:
Proceedings of the 1989 Jerusalem Winter School on Theoretical
Physics}, edited by S.Coleman, J.B.Hartle, T.Piran and S.Weinberg
(World Scientific, Singapore, 1991); and in ``Spacetime Quantum
Mechanics and the Quantum Mechanics of Spacetime'', preprint
UCSBTH92-91 (to appear in proceedings of the 1992 Les Houches Summer
School, {\it Gravitation et Quantifications}).

\refis{hartle2}
J. B. Hartle, ``The Reduction of the State Vector and
Limitations on Measurement in the Quantum Mechanics of Closed
Systems'',
in the Festschrift for D. Brill, edited by
B.L. Hu and T. Jacobson (Cambridge University Press, Cambridge, 1993).

\refis{hartle3} J.B.Hartle, \pr {\bf D44}, 3173 (1991)

\refis{grabert} H.Grabert, P.Schramm, G-L. Ingold, \prep {\bf 168},
115 (1988).

\refis{griffiths} R.Griffiths, \jsp {\bf 36}, 219 (1984).

\refis{hupaz} B.L.Hu, J.P.Paz and Y.Zhang, \pr {\bf D45}, 2843
(1992); ``Quantum Brownian Motion in a General Environment. II:
Non-Linear Coupling and Perturbative Approach'', Los Alamos Preprint
lA-UR-92-1367 (1992).

\refis{jjh} J.J.Halliwell, \pr {\bf D46}, 1610 (1992).

\refis{jjh2} J.J.Halliwell, ``Quantum-Mechanical Histories and
the Uncertainty Principle.
I. \ Information-Theoretic Inequalities'',
Imperial College Preprint IC 92-93/26. gr-qc 9304039 (1993).
To appear in {\sl Phys.Rev.D}.

\refis{jjh3} J.J.Halliwell, in preparation.

\refis{mensky1} M.Mensky, \pl {\bf A155}, 229 (1991).
% The action uncertainty principle in continuous quantum measurements.

\refis{omnes} R.Omn\`es, \rmp {\bf 64}, 339 (1992), and references
therein.

\refis{omnes2} R.Omn\`es, {\sl J.Stat.Phys.} {\bf 57}, 357 (1989).

\refis{wigner} See, for example, E.P.Wigner, in {\it Quantum Theory
and Measurement}, edited by J.A.Wheeler and W.H.Zurek (Princeton
University Press, Princeton, NJ, 1983).

\refis{wignerfn} For an extensive discussion of the Wigner function
and related phase space distribution functions, see
N.Balazs and B.K.Jennings, \prep {\bf 104}, 347 (1984);
M.Hillery, R.F.O'Connell, M.O.Scully and E.P.Wigner, \prep {\bf
106}, 121 (1984).

\refis{huzhang} B.L.Hu and Y.Zhang, ``Uncertainty Relation at Finite
Temperature'', University of Maryland Preprint (1992).

\refis{paz} J.P.Paz, S.Habib and W.Zurek, \pr {\bf D47}, 488 (1993).

\refis{unruh} W.G.Unruh and W.H.Zurek, \pr {\bf D40}, 1071 (1989).

\refis{vernon} R.P.Feynman and F.L.Vernon, \annp {\bf 24}, 118 (1963).

%\refis{zeh} E.Joos and H.D.Zeh, {\sl Z.Phys.} {\bf B59}, 223 (1985).

\refis{zurekpazhabib} W.Zurek, S.Habib and J.P.Paz, ``Coherent
States via Decoherence'', Los Alamos Preprint LA-UR-92-1642 (1992).

\refis {zurek2} W.H.Zurek, in {\it Frontiers of Non-Equilibrium
Statistical Mechanics}, edited by G.T.Moore and M.O.Scully (Plenum,
New York, 1986).

\refis {calzetta} E. Calzetta and B. L. Hu, ``Decoherence
of Correlation Histories'', to appear
in {\it Directions in General
Relativity}, edited by B. L. Hu and T. A. Jacobson (Cambridge
University Press, Cambridge, 1993).

\refis {diosi} L. Di\'osi,
``Unique Family of Consistent Histories in the
Caldeira-Leggett Model'',
Budapest preprint KFKI-RMKI-23 (1993).

\refis{isham} C. Isham,
``Quantum Logic and the Histories Approach to Quantum Theory'',
Imperial Preprint TP/92-93/39 (1993).

\refis{twamley} J. Twamley,
``Phase Space Deocherence: A Comparison between Consistent
Histories and Environment-Induced Superselection'',
Adelaide preprint ADP-93-208/M19 (1993).

\refis{saunders} S. Saunders,
``Decoherence, Relative States and Evolutionary Adaptation'',
Harvard Philosophy Department preprint (1993).

\endreferences
\end